\documentclass[aps,amssymb,amsmath,pra, twocolumn,superscriptaddress]{revtex4}
\usepackage{amsmath}
\usepackage{graphicx}
\usepackage{amsfonts}
\usepackage{epsfig}
\usepackage{hyperref}
\usepackage{amssymb}
\usepackage{bm}
\usepackage{bbm}
\usepackage{subfigure}
\usepackage{color}
\newcommand{\be}{\begin{equation}}
\newcommand{\ee}{\end{equation}}
\newcommand{\bc}{\begin{center}}
\newcommand{\ec}{\end{center}}
\newcommand{\bea}{\begin{eqnarray}}
\newcommand{\eea}{\end{eqnarray}}

\begin{document}
\title{Zeno subspace in quantum-walk dynamics}
\author{C. M. \surname{Chandrashekar}}
\email{chandru@imsc.res.in}
\affiliation{Quantum Science Center, The Institute of Mathematical Sciences, Chennai 600113, India}

\begin{abstract}
We investigate the discrete-time quantum-walk evolution under the influence of periodic measurements in position subspace. The undisturbed survival probability of the particle at the position subspace $P(0, t)$ is compared with the survival probability after frequent ($n$) measurements at interval $\tau = t/n$, $P(0, \tau)^n$.  We show that $P(0, \tau)^n > P(0, t)$ leads to the quantum Zeno effect in the position subspace when a parameter $\theta$  in the 
quantum coin operations and frequency of measurements is greater than the critical value, $\theta > \theta_{c}$ and $n>n_{c}$.  This Zeno effect in the subspace preserves the dynamics in coin Hilbert space of the walk dynamics and has the potential to play a significant role in quantum tasks such as preserving the quantum state of the particle at any particular position, and to understand the Zeno dynamics in a multidimensional system that is highly transient in nature. 
\end{abstract}

\maketitle
\preprint{Version}
\section{Introduction}
\label{intro}
In standard quantum theory, the time evolution of the state vector of the quantum system undergoes continuous unitary evolution until the system is measured. If very frequent measurements are performed on a quantum system, in order to ascertain whether it is still in its initial state, transitions to other states are hindered or boosted resulting in the quantum Zeno effect (QZE) or the inverse quantum Zeno effect (IZE), respectively \cite{BN67, Kha68, MS77, OPT93,  KS05, FP08}. QZE is expected to occur widely in quantum systems. In particular, for time $t$ with $n$ measurements, the complete suppression of the transition to other state in the limit of $t / n  \rightarrow 0$  is universal, common to all quantum systems, that is, the system is frozen to the initial state. However, in a multidimensional system the QZE does not necessarily freeze everything. On the contrary, for frequent projections onto a multidimensional subspace, the system can evolve away from its initial state, although it remains in the subspace defined by the measurement. This continuing time evolution within the projected subspace has also been investigated \cite{MS77, MNP99, FGM00, FPS02, FP02}. In this paper, the quantum system we use to investigate the Zeno dynamics in the projected subspace is the discrete-time quantum-walk (DTQW) evolution in $2 \times K$ Hilbert space.  
\par
Quantum-walk (QW) evolution involves the quantum features of interference and superposition resulting in the quadratically faster spread in position space than its classical counterpart, classical random walk (CRW)  \cite{Ria58, FH65, ADZ93,  DM96, FG98} in one dimension. QWs  are  studied  in   two  forms:
continuous-time  QW  (CTQW) \cite{FG98}  and  discrete-time QW  (DTQW)
\cite{ADZ93, DM96, ABN01, NV01} and are found to be very useful from the perspective of quantum algorithms \cite{Amb03, CCD03, SKB03, AKR05} (e.g., to demonstrate the coherent  quantum control over atoms, quantum phase transition \cite{CL08}; to explain phenomena such as the  breakdown of an electric-field driven  system \cite{OKA05} and direct  experimental  evidence  for  wavelike energy  transfer  within photosynthetic systems \cite{ECR07};  to generate entanglement between spatially separated systems \cite{CGS10}; and to induce Anderson localization of Bose-Einstein condensate in optical lattice \cite{Cha10b}). On the experimental front, implementation of QWs with samples in  an NMR system \cite{DLX03}; in the continuous tunneling of light fields  through waveguide lattices \cite{PLP08}; in  the  phase space  of trapped  ions  \cite{SMS09}; with single optically trapped  neutral atoms  \cite{KFC09}; and  with  single photon \cite{SCP10} has been reported. Various  other  schemes have  been proposed for  their physical  realization in different  physical systems \cite{RKB02, EMB05, Cha06}.
\par
Unlike many quantum processes on which the QZE is widely studied, DTQW is a controlled unitary evolution in which the constructive interference is directed away from the initial position $x=0$. This reduces the amplitude of the particle at $x=0$ to a very small value after the first few steps of the QW evolution (cf. recurrence nature of QW \cite{SJK08,  Cha08b}), thus, making the walk highly transient in nature. Introducing a decoherence channel to effectively mask the unitary evolution during each step of the DTQW decreases the transient behavior; therefore, the QZE can be shown by taking the rate of the measurement to $\infty$. (cf. Ref. \cite{AR05, VKB08} which discusses the QZE in CTQW). However, decoherence does no preserve the state subjected to QW evolution. In this paper we show that without introducing a decoherence channel, the unitary walk dynamics can be controlled to make it less transient by choosing the specific range of parameter in the quantum coin operation. Such a walk under the influence of periodic measurements in position subspace ($x=0$) is shown to lead to the QZE preserving the state of the particle at that position. This yields a quantum Zeno subspace in which the dynamics in the coin Hilbert space ${\cal H}_{c}$ of the walk is preserved. This observation can have implications for applications of QW to various quantum tasks such as preserving the quantum state \cite{DGR06} and quantum simulation of annealing processes \cite{SBB08}.
\par
This paper is arranged  as  follows. In  Sec.  \ref{dtqw}  we describe  the  DTQW  model on  a  line and its transient nature.   In  Sec. \ref{zenoeffect}  we discuss the conditions of the walk dynamics leading to the Zeno effect in the position subspace. Finally, in Sec. \ref{conc} we make our concluding remarks.

\section{Discrete-time quantum-walk and its transient nature}
\label{dtqw}
The DTQW  in  one-dimension is  modeled  as  a  $2\times K$ system, that is, a particle consisting  of  a two-level coin  (a qubit) in the  Hilbert space  ${\cal H}_c$, spanned  by $|0\rangle$  and  $|1\rangle$, and  $K$ positions in the position Hilbert  space  ${\cal  H}_p$,  spanned by $|\psi_x\rangle$, where $x \in {\mathbbm  I}$, the set of integers. A  $t$-step DTQW with unit time required for each step of walk is generated by  iteratively applying  a unitary  operation $W$  that acts  on the Hilbert space 
${\cal H}_c\otimes    {\cal     H}_p$,  
\be
|\Psi_t\rangle=W^t|\Psi_{ins}\rangle,
\ee  
where  $|\Psi_{ins}\rangle$ is the initial state of the particle at a particular position. We will choose a symmetric superposition state of the particle at position $x=0$,  
\be
|\Psi_{ins}\rangle= \frac{1}{\sqrt 2}(|0\rangle + i |1\rangle) \otimes |\psi_{0}\rangle
\ee
as initial state throughout this paper. $W\equiv S(B \otimes  {\mathbbm 1})$, where 
\be
\label{qwcoin}
B = B_{\xi,\theta,\zeta}       \equiv      \left(      \begin{array}{clcr}
  \mbox{~}e^{i\xi}\cos(\theta)      &     &     e^{i\zeta}\sin(\theta)
  \\ -e^{-i\zeta} \sin(\theta) & &  e^{-i\xi}\cos(\theta) 
\end{array} \right)\in SU(2)
\ee
is  the quantum coin operation.   $S$ is the controlled-shift operation          
 \be            S\equiv         \sum_x \left [  |0\rangle\langle
0|\otimes|\psi_x-1\rangle\langle   \psi_x|   +  |1\rangle\langle
1|\otimes |\psi_x+1\rangle\langle \psi_x| \right ]. 
 \ee
The probability to find the particle at position $x$ after $t$ steps is given by
\be
P(x,t)  = \langle
\psi_x|{\rm tr}_c (|\Psi_t\rangle\langle\Psi_t|)|\psi_x\rangle.
\ee
For a walk on a particle with the initial state at the origin $|\Psi_{ins} \rangle$
using an unbiased coin operation, that is, $B_{0, \theta, 0} \equiv B_{\theta}$), the variance after $t$ steps of walk is $[1 - \sin(\theta)] t^2$ and a symmetric probability distribution in position space is obtained \cite{CSL08}. 
\begin{figure}
\includegraphics[width=8.0cm]{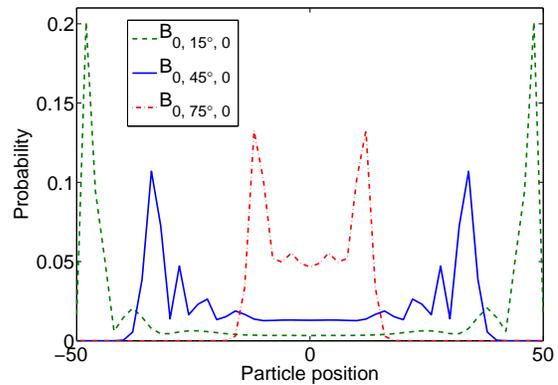}
\caption{\label{fig1}(Color online) The spread of probability distribution of the QW evolution using  different value of $\theta$ in the quantum coin operation $B_{\theta}$. The distribution is wider for  $\theta = 15^{\circ}$ and decreases with increase in $\theta$. Interference effect is absent for $\theta = 0^{\circ}$ and $\theta = 90^{\circ}$. The distribution is for 50 steps of the walk and only even positions are plotted because the odd position will have zero value for even number of steps.}
\end{figure}
In Fig. \ref{fig1}, probability distribution of 50-step QW evolution for different values of $\theta$ in the quantum coin operation $B_{\theta}$ is show.  For, $\theta=0^{\circ}$, the two states $|0\rangle$ and $|1\rangle$ move away from each other ballistically without any interference effect. With an increase in $\theta$ the interference effect is seen, and the distribution which is wider for low value of $\theta$ decreases with increase in $\theta$. The interference effect again disappears for another extreme value of $\theta = 90^{\circ}$. The two horned peaks on either side of the distribution, which moves away with an increase in the number of steps, makes QW highly transient in nature.
\par
\begin{figure}
\includegraphics[width=8.0cm]{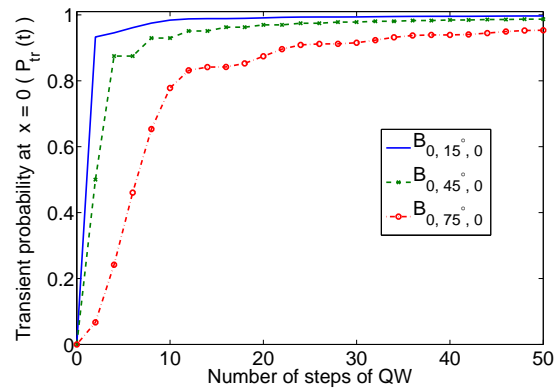}
\caption{\label{fig2}(Color online) Transient probability,  $P_{tr} (t) = [1 - P(0, t)]$,
probability of particle moving away from the initial position $x=0$ with number of steps (time) for QW with $\theta = 15^{\circ}$, $\theta = 45^{\circ}$ and $\theta = 75^{\circ}$. We note that the  $P_{tr}(t)$ shoots up very quickly for lower value of $theta$ and with increase in $\theta$, $P_{tr}(t)$ increases gradually.}
\end{figure}
In Fig. \ref{fig2} we have plotted the transient probability  
\be
P_{tr} (t) = [1 - P(0, t)],
\ee
that is, the probability of particle moving away from the initial position $x=0$ with number of steps (time). For lower value of $\theta$, the  $P_{tr}(t)$ shoots up very quickly, and with an increase in $\theta$, $P_{tr}(t)$ increases gradually. Therefore, making a measurement at position $x=0$ for large value of $\theta$ will yield a survival probability.  
This behavior is the key for us to explore the quantum Zeno region (QZR) in the DTQW evolution. 
\section{Zeno effect in subspace of discrete-time quantum-walk}
\label{zenoeffect}
In this section we outline the conditions on performing the measurements and using quantum coin parameters to observe the QZE in the subspace of the walk dynamics, preserving the dynamics in the coin Hilbert space ${\cal H}_c$. We first consider the position $x=0$ as the subspace, ${\cal H}^s_p \in {\cal H}_p$ from the complete Hilbert space of the DTQW system ${\cal H} = {\cal H}_{c} \otimes {\cal H}_p$ to study the QZE. One of the most trivial way to freeze the particle at subspace ${\cal H}^s_p$ with $P(0, t) =1$ resulting in the QZE, is by making projective measurements in ${\cal H}^s_p$ at intervals far less than the time required to implement one step of the walk ($\tau < <1$, with unit time required for each step). 
Due to transient nature of the DTQW, for $\tau \geq 1$, $P(0, \tau) \neq 1$, observing the Zeno effect is not straightforward. 
However, by being selective in performing measurements, we can see the QZR \cite{FP98}  if the undisturbed survival probability of the particle in ${\cal H}^s_p$ is less than the survival probability with $n$ measurements, that is,  
\be
P(0, t) < P(0, \tau)^{n},
\ee
where $t = n\tau$. 
Measurements has to be selective for DTQW evolution because, for every odd number of steps of walk, the probability at subspace ${\cal H}^s_p$ and other even positions is always zero. Therefore, intervals we need to consider for measurements are $t=2$ and its multiples. If we consider the measurement after the first two step of the walk, the state only at subspace ${\cal H}^s_p$ is retained and the rest is discarded, the QW is further evolved for next two steps, and the process is repeated many times before calculating the survival probability. 
\par
\begin{figure}
\includegraphics[width=8.5cm]{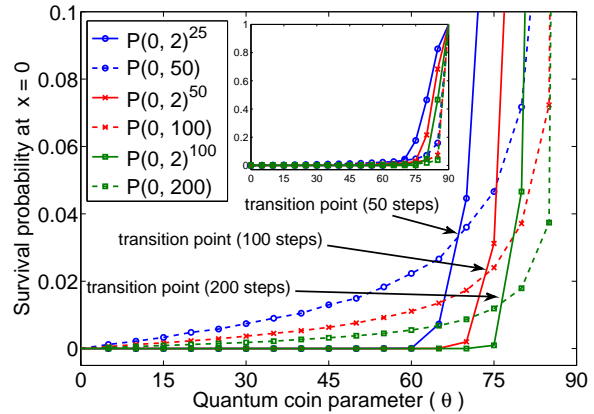}
\caption{\label{fig3}(Color online) Closer view of the variation of the survival probability at position $x = 0$ with $\theta$ for 50, 100 and 200 steps QW evolution with measurement after every two steps (solid lines) and after one single measurement at the end of the evolution (dashed line).   Transition of the dynamics to the quantum Zeno region, $P(0, 2)^{n} > P(0, 2n)$ is observed at different $\theta$ for walk with different number of steps. Inset is a full plot of survival probability becoming a unit value at $\theta = \pi/2$.}
\end{figure}
It is convenient to discuss the evolution of $t$-step walk using $W\equiv S(B \otimes  {\mathbbm 1})$ in terms of  the density matrix $\rho(x, t)$. If $\rho(0,0) = |\Psi_{ins}\rangle \langle \Psi_{ins}|$, density matrix after $t$ steps of walk will be
\bea
\rho(x,t) = (W^t) \rho(0, 0) (W^t) ^{\dagger}.
\eea
By taking projective measurement on the position subspace $|\psi_{0}\rangle$, we get
\bea
\rho(0, t) = \langle \psi_{0} | (W^t) \rho(0, 0) (W^t) ^{\dagger} |\psi_{0} \rangle.
\eea
Then the survival probability at subspace ${\cal H}^s_p$ after a single measurement is
\be
P(0, t) =  \mbox{tr}_c [\rho(0, t)].
\ee
If the projective measurement on the position subspace ${\cal H}^s_p$  is made after the first two steps of the walk, the state $| \Psi(0, 2)\rangle$ will be
\bea
|\Psi(0, 2)\rangle = [-e^{i(\xi - \zeta)}\cos(\theta)\sin(\theta) - i \sin^{2}(\theta)]|0\rangle \nonumber 
\\ 
+  [-\sin^{2}(\theta) + i e^{-i(\xi - \zeta)}\cos(\theta)\sin(\theta)]|1\rangle.
\eea
In the preceding expression, we can note that the survival probability is largely dependent on $\theta$ and, due to symmetric contribution from the neighboring lattice to the position $x=0$, one can ignore the role of $\xi$ and $\zeta$ in the survival probability of the state.  The  general form of the two-component vector of amplitudes of the particle, being at position $x$ and at  time $t$,  with left-moving ($L$) and right-moving ($R$) components, can be written in the form 
\begin{eqnarray}
\label{eq:compa}
\left ( \begin{array}{cl} \Psi_L(x,t)\\
\\
\\
\Psi_R(x,t) 
\end{array}
\right )  = \left (\begin{array}{cl} e^{i\xi}\cos(\theta)\Psi_L(x+1,t-1) \\
	+ e^{i\zeta} \sin(\theta)\Psi_R(x-1,t-1)   \\
 \\
 e^{-i\xi}\cos(\theta)\Psi_R(x-1,t-1)  \\
 - e^{-i\zeta} \sin(\theta)\Psi_L(x+1,t-1)
\end{array}
\right ).
\end{eqnarray}
For $x=0$ and any time $t$  in the preceding expression [the argument used for evolution with measurement after two step, Eq. (\ref{eq:compa})], the symmetric contribution from the neighboring lattice remains valid. 
\par
When $n$ periodic measurements are made on the system,
\be 
P(0, \tau)^n = \mbox{tr}_{c}[(W^{\tau}_{M})^{n} \rho(0, 0){(W^{\tau}_{M})^{n}}^{\dagger}] \equiv \mbox{tr}_{c}[\rho(0, \tau)^{n}],
\ee
where $W^{\tau}_{M}$ is the unitary operation $W$ with projective measurement onto ${\cal H}^{s}_{P}$ after $\tau$ operations.
\begin{figure}
\includegraphics[width=7.8cm]{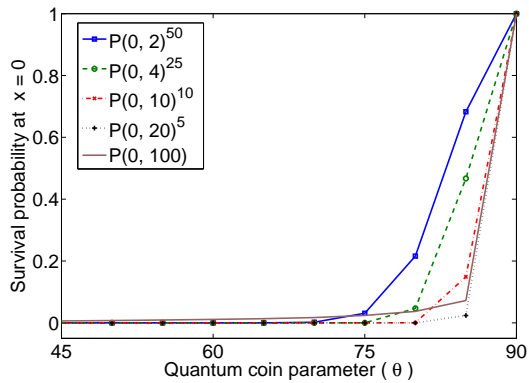}
\caption{\label{fig4}(Color online) Variation of survival probability with $\theta$ for 100 steps QW evolution with different frequency of measurements. With increase in frequency of measurement, the QZR for the a range of $\theta$ increases.}
\end{figure}
\begin{figure}
\includegraphics[width=7.8cm]{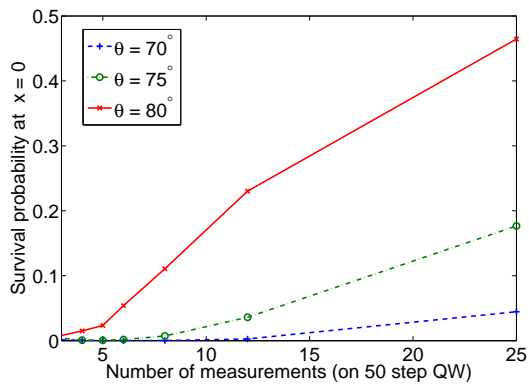}
\caption{\label{fig5}(Color online) Variation of survival probability with number of measurements (n) for 50 steps of QW evolution with different $\theta$. With increase in frequency of measurement and $\theta$, the QZR increases. All the three plots in the figure converges to unit value for n (x -axis) = 50.}
\end{figure}
For a Hadamard walk it is shown that $P(0, t) = O(t^{-1})$ \cite{NV01} and for walk with coin operation $B$,  $P(0, t) = O( \frac{\sin(\theta)}{t})$.  Therefore, $P(0, \tau)^n = O ( \frac{\sin(\theta)}{\tau})^n$. For larger $\theta$ and small $\tau$ (i.e., more frequent measurement), we can see that $P(0, \tau)^n > P(0,t)$ leading to QZR.
\par
In Fig. \ref{fig3}, undisturbed and disturbed survival probability of the state of the particle at position $x = 0$ for QW evolution with different $\theta$ in the quantum coin operation are shown for 
50, 100 and 200 steps of walk. The undisturbed survival probability, $P(0, t)$  is shown using dashed lines, and the disturbed survival probability with $n$ measurements after every two steps, $P(0, 2)^n$, is shown using solid lines. For a QW with measurements after every 2 steps, $P(0,2)^{25} \approx 0$ for $\theta<60^{\circ}$ (for 50 steps), $P(0,2)^{50} \approx 0$ for $\theta<65^{\circ}$ (for 100 steps), and $P(0,2)^{100} \approx 0$ for $\theta<70^{\circ}$ (for 200 steps), respectively. 
At the transition point leading to the QZR, $P(0, \tau)^{n}  = P(0, n\tau)$.  Beyond $\theta > \theta_{c}$ we note that $P(0, 2)^{n} < P(0, 2\times n)$, leading the walk dynamics to  the QZR. 
In Fig. \ref{fig4}, survival probability with $\theta$ for 100-step QW evolution with different frequencies of measurements is shown. The QZR increases with increase in the frequency of measurements for a range of $\theta$. In Fig. \ref{fig5},  survival probability with number of measurements for 50-step of QW evolution with different $\theta$ is shown. With an increase in the frequency of measurement and $\theta$, the QZR also increases. 
For lower value of $\theta$ in Fig. \ref{fig3}, even though $P(0, \tau)^n < P(0, n \tau)$, we note that it does not continue to decrease (i.e., opposite of Fig. \ref{fig5}) with an increase in the number of measurements. This suggests the absence of Inverse-QZR.
\section{concluding remarks}
\label{conc}
 We have discussed the DTQW evolution under the influence of the periodic measurements in position subspace that yields Zeno subspace preserving the dynamics of the coin Hilbert space. The transient nature of the DTQW, which decreases with an increase in the value of parameter $\theta$ in the quantum coin operation, was used to explore the QZR.  For particular value of $\theta$ and frequency of measurements, the transition from 
survival probability with measurement less than the undisturbed survival probability, to survival probability with measurement greater than the undisturbed survival probability $P(0, t)< P(0, \tau)^n$ when $t=n\tau$, was shown leading the transition to the QZR. Because we did not consider the decoherence channel to suppress the walk dynamics leading to Zeno effect, the dynamics of the state of the particle in the projected subspace is preserved. These observations can have implications for  applications of DTQW to various quantum tasks. In Ref. \cite{DGR06}, an algorithm to preserve the quantum state was proposed making use of the QZE. Using DTQW and QZE with periodic measurements in position subspace can also be used to preserve the quantum state of the particle, not only at $x=0$ but also at any other particular position in the position space. This can be achieved by using the combination of undisturbed DTQW evolution, first to shift the peak of the QW to desired location, followed by frequent measurements using the parameters that can result in QZE. In Ref. \cite{SBB08}, quantum simulation of the classical annealing system is proposed using the combination of the CTQW and the QZE in its dynamics. The ability to control the DTQW dynamics and the QZE using quantum operations can lead to further exploration of annealing problem.
\par
{\it Acknowledgement --} I thank Prof. R. Simon for encouragement.

\end{document}